\documentclass[12pt]{article}
\usepackage{epsfig,amsfonts,amssymb}
\usepackage{hyperref}
\usepackage{cite}
\input epsf.sty
\usepackage{cite}

\def\ZZZ{{\hbox{ Z\kern-1.6mm Z}}}
\def\RRR{{\hbox{ R\kern-2.4mm R}}}
\def\CCC{{\hbox{ C\kern-2.0mm C}}}
\def\zzz{{\hbox{z\kern-1mm z}}}

\newcommand{\nn}{\nonumber \\}

\newcommand{\qeq}{{\hbox{=\kern-2.3mm ? \kern.5mm }}}
\renewcommand{\qeq}{=}

\newcommand{\DD}{{\cal D}}

\newcommand{\LL}{{\cal L}}

\newcommand{\NN}{{\cal N}}

\newcommand{\SSS}{{\cal S}}

\newcommand{\be}{\begin{equation}}
\newcommand{\ee}{\end{equation}}
\newcommand{\ben}{\begin{eqnarray}\displaystyle}
\newcommand{\een}{\end{eqnarray}}

\newcommand{\refb}[1]{(\ref{#1})}
\newcommand{\p}{\partial}
\newcommand{\sectiono}[1]{\section{#1}\setcounter{equation}{0}}

\def\one{{\hbox{ 1\kern-.8mm l}}}
\def\zero{{\hbox{ 0\kern-1.5mm 0}}}

\newcommand{\bea}[1]{\begin{eqnarray}\label{#1} }
\newcommand{\eea}{\end{eqnarray}}

\begin{document}

\baselineskip 24pt

\begin{center}
{\Large \bf  Heat Kernel Expansion and Extremal Kerr-Newmann
Black Hole
Entropy in Einstein-Maxwell Theory}

\end{center}

\vskip .6cm
\medskip

\vspace*{4.0ex}

\baselineskip=18pt

\centerline{\large \rm Sayantani Bhattacharyya, Binata Panda and Ashoke Sen}

\vspace*{4.0ex}

\centerline{\large \it Harish-Chandra Research Institute}
\centerline{\large \it  Chhatnag Road, Jhusi,
Allahabad 211019, India}

\vspace*{1.0ex}
\centerline{\small E-mail:  sayanta,binatapanda,sen@mri.ernet.in}

\vspace*{5.0ex}

\centerline{\bf Abstract} \bigskip

We compute the second Seely-DeWitt  coefficient of the kinetic
operator of the metric and gauge fields in Einstein-Maxwell theory
in an arbitrary background field configuration. 
We then use this result to compute the logarithmic
correction to the entropy of an extremal Kerr-Newmann black hole.

\vfill \eject

\baselineskip=18pt

\section{Introduction} \label{s1}

In a classical two derivative theory of gravity, the Bekenstein-Hawking formula
gives an expression for the entropy of the black hole. This formula is
universal in the sense that it does not depend on the details of the matter
field content of the theory. Furthermore it does not depend on any specific
ultraviolet completion of the theory -- an independent computation of the
black hole entropy from counting of microstates must reproduce this result
in any consistent quantum theory of gravity.
This provides a strong constraint on consistent ultraviolet completions 
of the theory.

The Bekenstein-Hawking result for the entropy is expected to
receive quantum
corrections which are subdominant for black holes of large size but
could become important for black holes of size comparable to Planck size. The
details of these corrections will depend on the ultraviolet completion of the
theory, but there is a certain class of corrections, proportional to the logarithm
of the size of the black hole, which receive contribution from loops of massless
fields propagating in the black hole background and 
can be computed purely from the knowledge
of the infrared 
physics\cite{9407001,9408068,9412161,9604118,9709064,
1005.3044,1106.0080,1108.3842,1109.3706,1109.0444,1008.4314,1104.3712}.
In particular for a class of extremal black holes in four dimensional $\NN=4$
and $\NN=8$ supersymmetric string theories and five dimensional
BMPV black holes\cite{9601029,9602065} the logarithmic corrections to the
black hole entropy computed from infrared physics correctly reproduce the
result from the microscopic counting in string
theory\cite{1005.3044,1106.0080,1109.3706}.

Since the procedure used in \cite{1005.3044,1106.0080,1108.3842,1109.3706}
for computing logarithmic correction to the entropy
does not rely on supersymmetry, this can be carried out for any extremal 
black hole. This requires computing the heat kernel (more precisely its
expansion coefficient $a_4$ introduced in eq.\refb{esmalls}) 
of the kinetic operator of the 
massless fields in the black
hole background. Keeping this motivation in view we study the short distance
expansion of the heat kernel in Maxwell-Einstein theory for fluctuations
of gauge and graviton fields around an arbitrary background field configuration
and compute
the relevant quantity $a_4$ that appears as the coefficient of the
logarithmic corrrection to the black hole entropy. The result is given
in eq.\refb{ea4fin}. We then evaluate this in the
background of an extremal Kerr-Newmann black hole solution and find
an explicit expression
for the coefficient of the logarithmic correction to the black hole entropy. 
The result is in eq.\refb{elogfin}. Any
microscopic explanation of the entropy of such a black hole,
{\it e.g.} via Kerr-CFT correspondence\cite{0809.4266,1203.3561}, 
should reproduce the result 
given
in \refb{elogfin}.

\sectiono{Heat kernel expansion in Einstein-Maxwell theory} \label{s2}

We consider Euclidean continuation of
Einstein-Maxwell theory in four space-time dimensions with
action
\be \label{ernk1}
\SSS = \int d^4 x\, \sqrt{\det g}\,  \LL, \qquad
\LL= \left[ R - F_{\mu\nu} F^{\mu\nu}\right]\, ,
\ee
where $R$ is the scalar curvature computed with
the metric $g_{\mu\nu}$ and $F_{\mu\nu}
=\p_\mu A_\nu - \p_\nu A_\mu$ is the gauge field
strength. 
Note that we have set $G_N = 1/16\pi$. Let $(\bar g_{\mu\nu}, \bar A_\mu)$
denote any solution to the classical equations of motion in this theory and
let $\bar F_{\mu\nu}\equiv \p_\mu \bar A_\nu - \p_\nu \bar A_\mu$ be the
corresponding gauge field strength. We consider fluctuations around this
background of the form
\be \label{ernk2}
g_{\mu\nu} = \bar g_{\mu\nu} + \sqrt 2 \, h_{\mu\nu}, \qquad
A_\mu = \bar A_\mu + {1\over 2}\, 
a_\mu \, ,
\ee
and denote by $\{\phi_m\}$ the set of all the fluctuating fields
$\{h_{\mu\nu}, a_\mu\}$.
Then to quadratic order in the fluctuations the action takes the 
form:
\be \label{equad}
{1\over 2} \int d^4 x \sqrt{\det \bar g} \, M^{mp}
 \phi_p K_{m}^{~n} \phi_n\, ,
\ee
for some matrix $M$ and a differential operator $K$
of the form
\be \label{ekform}
K_m^{~n} = (D^\mu D_\mu)_m^{~n} + (N^\mu D_\mu)_m^{~n} 
+ P_m^{~n}\, ,
\ee
where $D_\mu$ denotes ordinary covariant derivative with connections
determined by the background metric and
$N^\mu$ and $P$ are appropriate matrices constructed from the
background fields. All indices are raised
and lowered by the background metric $\bar g_{\mu\nu}$, and
the total derivative terms are adjusted so that the operator $K$
is hermitian. 
Then the heat kernel is defined to be the operator $e^{s \, K}$,
and 
${\rm Tr} \, e^{s \, K}$ has a small $s$ expansion of the 
form\cite{sd,duffobs,christ-duff1,christ-duff2,duffnieu,birrel,gilkey,
0306138,1009.4439}:
\be \label{esmalls}
{\rm Tr} \, e^{s\, K} = \int d^4 x \, \sqrt{\det \bar g} \, \sum_{n=0}^\infty 
s^{n-2} a_{2n}(x)\, ,
\ee
where the coefficients $a_{2n}(x)$ -- known as the Seely-DeWitt 
coefficients\cite{sd} -- are expressed in terms of local
invariants constructed from the background metric, 
Riemann tensor, gauge field strength and their
covariant derivatives. Our goal will be to compute the coefficient
$a_4$ for an arbitrary background  $(\bar g_{\mu\nu}, \bar A_\mu)$
satisfying classical equations of motion,
and then use this to compute logarithmic corrections to the entropy
of an extremal Kerr-Newmann black hole using the results of
\cite{1005.3044,1106.0080,1108.3842,1109.3706}.\footnote{On manifolds
with boundary we can also have boundary tems with half integral
powers of $s$.}

Our strategy for computing $a_4$ is as follows. First we express
the differential operator $K_m^{~n}$ as
\be \label{enform}
K_m^{~n}= (\DD_\mu \DD^\mu)_m^{~n} + E_m^{~n}\, ,
\ee
where 
\ben \label{edefddetc}
\DD_\mu &=& D_\mu + \omega_\mu, \quad \omega_\mu = {1\over 2}
\bar g_{\mu\nu} N^\nu, \nn
E &=& P - \bar g^{\mu\nu} (\omega_\mu \omega_\nu + D_\mu \omega_\nu)
\, .
\een
If 
\be \label{edefom}
\Omega_{\mu\nu}\equiv [\DD_\mu,\DD_\nu]\, ,
\ee 
denotes the curvature associated
with the covariant derivative $\DD_\mu$,
then we have\cite{0306138}
\ben \label{ea4coeff}
a_4(x) &=& {1\over 360 \times 16\pi^2} {\rm tr}\, \bigg[60 D^\mu D_\mu E + 60 R E
+ 180 E^2 + 12 D_\mu D^\mu R \nn
&& \qquad \qquad \qquad + 5 R^2 - 2 R_{\mu\nu} R^{\mu\nu}
+ 2 R_{\mu\nu\rho\sigma} R^{\mu\nu\rho\sigma} + 30
\Omega_{\mu\nu} \Omega^{\mu\nu} 
\bigg]\, .
\een
Here $R_{\mu\nu\rho\sigma}$, $R_{\mu\nu}$ and $R$ denote 
respectively the
Riemann tensor, Ricci tensor and scalar curvature computed from the
background metric and tr denotes trace over the index $m$
labelling all the
fluctuating fields $\phi_m$. In our analysis we shall ignore total derivative
terms which do not contribute to the integral in \refb{esmalls}. Also
for any classical solution we have $R=0$ and hence we can ignore terms
proportional to $R$. 

We gauge fix the Einstein-Maxwell theory by adding
a gauge fixing term
\be \label{egaugefix}
-\int d^4 x \, \sqrt{\det \bar g}\, \left[ 
g^{\rho\sigma}\,
\left(D^\mu h_{\mu\rho} -{1\over 2} D_\rho \, 
h^\mu_{~\mu}\right)
\left( D^\nu \, h_{\nu\sigma} -{1\over 2} 
D_\sigma h^\nu_{~\nu}\right)
 + {1\over 2} D^\mu a_\mu D^\nu a_\nu\right]
\, ,
\ee
and add the corresponding ghost action.
The total gauge fixed action up to quadratic order
in the fluctuations is given by
\ben \label{egfaction}
\SSS &=& {1\over 2} \, \int d^4 x\, \sqrt{\det \bar g}\, \bigg[
-h^{\mu\nu} (\Delta h)_{\mu\nu} + a_\mu \left\{ (D_\rho D^\rho)^{\mu\nu} -
R^{\mu\nu}\right\} a_\nu \nn && 
- {1\over 2} \bar F_{\mu\nu} \bar F^{\mu\nu}  \left\{
(h^\rho_{~\rho})^2 - 2 h^{\rho\sigma} h_{\rho\sigma}\right\}
- 4 \bar F_{\mu\nu} \bar F_{\rho\sigma} h^{\mu\rho} h^{\nu\sigma}
- 8 \bar F_{\mu\rho} \bar F_{\nu}^{~\rho} h^\mu_{~\sigma} h^{\nu\sigma}
+ 4 \bar F_{\mu\rho} \bar F_{\nu}^{~\rho} 
h^\sigma_{~\sigma} h^{\mu\nu} \nn &&
- \sqrt 2 \bar F^{\mu\nu} h^\sigma_{~\sigma} f_{\mu\nu}
+ 4\sqrt 2\, \bar F^{\mu\nu} h_\nu^{~\rho}  f_{\mu\rho}
\, \nn &&
+ 2 b^\mu \left(\bar g_{\mu\nu}\square + R_{\mu\nu}\right) 
 c^\nu + 2 b\square c
- 4 \, b \bar F_{\mu\nu} \, D^\mu c^\nu
\bigg]\, , \nn
\een
where 
\be \label{edeffmunu}
f_{\mu\nu}\equiv \p_\mu a_\nu - \p_\nu a_\mu\, ,
\ee
\ben \label{e35}
\left(\Delta h\right)_{\mu\nu} 
&=& -\square h_{\mu\nu} - R_{\mu\tau} h^\tau_{~\nu}
- R_{\nu\tau} h_\mu^{~\tau} - 2 R_{\mu\rho\nu\tau} h^{\rho\tau}
+{1\over 2} \, \bar g_{\mu\nu} \, 
\bar g^{\rho\sigma} \, \square\, h_{\rho\sigma}
\nn && + R\, h_{\mu\nu} 
+ \left(\bar g_{\mu\nu} R^{\rho\sigma} + R_{\mu\nu}
\bar g^{\rho\sigma}\right) h_{\rho\sigma} -{1\over 2}\, R\, 
\bar g_{\mu\nu} \, \bar g^{\rho\sigma}\, h_{\rho\sigma}
\, .
\een
$b_\mu, c_\nu$ are the diffeomorphism ghosts and $b,c$ are the ghosts
associated with the gauge invariance of the Maxwell action.
Note that the ghosts will contribute to the trace in
\refb{ea4coeff} with an overall negative sign.

We can determine the
matrices $M$, $N$, $P$
by comparing \refb{egfaction} with \refb{equad}, \refb{ekform} 
and then 
determine $\omega_\mu$, $E$ and $\Omega_{\mu\nu}$
using \refb{edefddetc},
\refb{edefom}.
This in turn allows us to compute $a_4$ via \refb{ea4coeff} in terms of
the background fields. We can simplify the expression using the
equations of motion:
\be \label{esimple}
D^\mu \bar F_{\mu\nu}=0, \qquad R_{\mu\nu} = 2 \bar F_{\mu\rho}
\bar F_\nu^{~\rho} - {1\over 2}  \bar g_{\mu\nu} \bar F_{\rho\sigma}
\bar F^{\rho\sigma}\, ,
\ee
 and the Bianchi identities:
 \be \label{ebianchi}
 D_{[\mu} \bar F_{\nu\rho]}=0, \qquad R_{\mu[\nu\rho\sigma]} = 0\, .
\ee
The computation is tedious but straightfiorward and so we shall not
give the details of the intermediate steps.
Up to addition of total derivative terms
the non-vanishing contribution to $a_4$ (including the
contribution from the ghosts) comes from the following terms:
\ben \label{eterms1}
&& {\rm tr} \left(- 2 R_{\mu\nu} R^{\mu\nu}
+ 2 R_{\mu\nu\rho\sigma} R^{\mu\nu\rho\sigma}\right)
= 4 \left(- 2 R_{\mu\nu} R^{\mu\nu}
+ 2 R_{\mu\nu\rho\sigma} R^{\mu\nu\rho\sigma}\right)\nn
&& {\rm tr} \, E^2 = 3 R_{\mu\nu\rho\sigma} R^{\mu\nu\rho\sigma}
- 9 R_{\mu\nu} R^{\mu\nu} + 3 R_{\mu\nu\rho\sigma}
\bar F^{\mu\nu} \bar F^{\rho\sigma} 
+ 9 \left(\bar F^{\mu\nu} \bar F_{\mu\nu}\right)^2 \nn 
&& {\rm tr} \, \left( \Omega_{\mu\nu} \Omega^{\mu\nu}\right)
= -5 R_{\mu\nu\rho\sigma} R^{\mu\nu\rho\sigma}
+ 56 R_{\mu\nu} R^{\mu\nu} -18 R_{\mu\nu\rho\sigma}
\bar F^{\mu\nu} \bar F^{\rho\sigma} 
-54 \left(\bar F^{\mu\nu} \bar F_{\mu\nu}\right)^2\, . \nn
\een
Substituting these in \refb{ea4coeff} we get
\be \label{ea4fin}
a_4(x) = {1\over 360 \times 16\pi^2} 
\left(398 R_{\mu\nu\rho\sigma} R^{\mu\nu\rho\sigma}
+ 52 R_{\mu\nu} R^{\mu\nu}\right)\, .
\ee
This will be one of our central results.

Before we proceed we note several points:
\begin{enumerate}
\item The terms proportional to $R_{\mu\nu\rho\sigma}
\bar F^{\mu\nu} \bar F^{\rho\sigma}$ and 
$\left(\bar F^{\mu\nu} \bar F_{\mu\nu}\right)^2$ cancel in the final
expression. Since the final result is given only in terms of the
background metric, the result is invariant under electric-magnetic
duality rotation.
\item For the euclidean
near horizon geometry of a Reissner-Nordstrom black hole:
\be \label{ernk1.5}
ds^2 \equiv \bar g_{\mu\nu} dx^\mu dx^\nu =
\Lambda^2(d\eta^2 +\sinh^2\eta d\theta^2) + \Lambda^2 (d\psi^2 +
\sin^2 \psi d\phi^2), \qquad \bar F_{\psi\phi}
= {i} \, \Lambda\, \sin\psi\, ,
\ee
eq.\refb{ea4fin} gives $a_4 = 53 / (90 \pi^2 \Lambda^4)$. This agrees with the
explicit computation of $a_4$ using the eigenvalues of the
kinetic operator $K$\cite{1108.3842}.
\item Even though the final result is given in terms of the background metric
only, we shall get wrong result if we ignore the terms involving
$\bar F_{\mu\nu}$ from the beginning. Part of the contribution involving
$R_{\mu\nu} R^{\mu\nu}$ term in \refb{ea4fin} comes as a result of
eliminating the combination $\bar F_{\mu\rho}
\bar F_\nu^{~\rho}$ in terms of $R_{\mu\nu}$ using \refb{esimple}.
\end{enumerate}

\sectiono{Logarithmic correction to Kerr-Newmann black hole entropy} \label{s3}

Next we shall use \refb{ea4fin} to compute logarithmic correction to the
entropy of an extremal Kerr-Newmann black hole. 
The metric of a 
general Kerr-Newmann black hole  is given by
\ben \label{ekerrmetric}
ds^2 &=& - {r^2 + a^2 \cos^2 \psi - 2 Mr +Q^2\over r^2 + a^2 \cos^2 \psi} dt^2
+ { r^2 + a^2 \cos^2 \psi   \over   r^2 + a^2  - 2 Mr +Q^2} dr^2 
+(r^2 + a^2 \cos^2 \psi) d\psi^2 \nn &&
+
{ (r^2 + a^2 \cos^2 \psi) (r^2+a^2) + (2 Mr -Q^2) a^2\sin^2\psi
\over  r^2 + a^2 \cos^2 \psi} \sin^2\psi d\phi^2 
\nn &&
+{2 (Q^2-2 M r) a\over r^2 + a^2 \cos^2 \psi} \sin^2\psi \, dt d\phi \nn 
&=& - {(r^2 + a^2 \cos^2 \psi) (r^2 + a^2  - 2 Mr +Q^2)\over
(r^2 + a^2 \cos^2 \psi) (r^2+a^2) + (2 Mr -Q^2) a^2\sin^2\psi} dt^2
\nn &&
+ { r^2 + a^2 \cos^2 \psi   \over   r^2 + a^2  - 2 Mr +Q^2} dr^2 
+(r^2 + a^2 \cos^2 \psi) d\psi^2 \nn &&
+ { (r^2 + a^2 \cos^2 \psi) (r^2+a^2) + (2 Mr -Q^2) a^2\sin^2\psi
\over  r^2 + a^2 \cos^2 \psi} \sin^2\psi \nn && 
\times \left(d\phi + { (Q^2-2 M r) a\over
(r^2 + a^2 \cos^2 \psi) (r^2+a^2) + (2 Mr -Q^2) a^2\sin^2\psi} dt
\right)^2\, ,
\een
where for $G_N=1/16\pi$
the physical mass $m$, charge $q$ 
and the angular momentum $J$ are related
to the parameters $M$, $Q$ and $a$ via
\be \label{ephysmj}
m = 16\pi M, \qquad J = 16 \pi M a, \qquad q = 8\pi \, Q\, .
\ee
The horizon is located at
\be \label{ehorloc}
r^2 - 2Mr + a^2 + Q^2 = 0\, ,
\ee
and the classical Bekenstein-Hawking entropy, given by 
$1/4G_N=4\pi$ times the area of the outer event horizon,
takes the form
\be \label{entropy}
S_{BH} = 16\pi^2 \left[2M^2 - Q^2 +2M\sqrt{M^2 - (a^2 + Q^2)}\right]\, .
\ee

The extremal limit corresponds to $M\to \sqrt{a^2+Q^2}$. In this limit
we get
\be \label{extremal}
S_{BH}=16\pi^2 (2a^2+Q^2) = \sqrt{(q^2/4)^2 + (2\pi J)^2}\, .
\ee
To take extremal limit of the near horizon geometry
we introduce a new parameter $\lambda$
and new coordinates $\rho,\tau,\chi$ via (see {\it e.g.} \cite{1008.3801}):
\be \label{enewcor}
M^2 = a^2 + Q^2 + \lambda^2, \quad r = M + \lambda\rho, \quad
t =(2a^2 + Q^2) \tau/\lambda, \quad \phi=\chi + {a\over Q^2 + 2 a^2}
\left(1 - {2M\lambda\over Q^2 + 2 a^2}\right) t \, ,
\ee
and take the $\lambda\to 0$ limit keeping $\rho$, $\tau$ fixed. In this limit
the metric \refb{ekerrmetric} reduces to
\ben \label{emetricred}
ds^2 &=& (Q^2 + a^2 + a^2 \cos^2\psi) \left( - (\rho^2-1) d\tau^2 
+ {d\rho^2\over
(\rho^2-1)} + d\psi^2\right) \nn
&& + {(Q^2+2a^2)^2\over(Q^2 + a^2 + a^2 \cos^2\psi)}
\sin^2\psi \left(d\chi +{2  M a\over Q^2 + 2 a^2}
(\rho-1) d\tau\right)^2\, .
\een
Finally after euclidean continuation and
another change of variables,
\be \label{eeucrot}
\tau=-i\theta, \qquad \rho=\cosh\eta\, ,
\ee
with $\theta$ interpreted as a periodic coordinate with period $2\pi$,
we get
\ben \label{emetricfin}
ds^2 &\equiv& \bar g_{\mu\nu} dx^\mu dx^\nu \nn
&=& (Q^2 + a^2 + a^2 \cos^2\psi) \left( d\eta^2 + \sinh^2\eta\, 
d\theta^2+
d\psi^2\right) \nn
&& + {(Q^2+2a^2)^2\over(Q^2 + a^2 + a^2 \cos^2\psi)}\, \sin^2\psi\,
 \left(d\chi -i{2  M a\over Q^2 + 2 a^2}
(\cosh\eta-1) d\theta\right)^2\, .
\een

Let us now consider the limit in which the charge and the angular momenta
become large as
\be \label{escaling}
q\sim \Lambda, \quad J\sim \Lambda^2 \, ,
\ee
for some large number $\Lambda$. In this limit $Q,a\sim\Lambda$
and the horizon area $A_H$ scales as
$\Lambda^2$. It follows from the analysis
of \cite{1109.3706} that in this limit the one loop quantum correction to the
entropy has a term proportional to $\ln \Lambda$, given by
\be \label{elogcoeff}
 \ln \Lambda \, \left[\sum_r (\beta_r-1) N_r  -2\pi \, \int_{\rm horizon} 
 d\psi d\phi \, G(\psi) \, a_4(x)
\right]\, ,
\ee
where
\be \label{edefgy}
G(\psi)=\sqrt{\det \bar g} /\sinh\eta 
= (Q^2+a^2 + a^2\cos^2\psi) (Q^2+2a^2) \sin\psi\, ,
\ee
and $\sum_r (\beta_r-1)N_r$ is the zero mode contribution evaluated as follows.
The sum over $r$ represents the sum over various fields, -- in this case the metric
and the U(1) gauge field. $\beta_r$'s are constants defined so that the Jacobian of
changing variables from integration over the fields to integration over the zero mode
deformation parameters gives a factor of $\Lambda^{\beta_r}$ per zero mode. In particular
$\beta_r=1$ for the four dimensional 
gauge fields and 2 for the four dimensional metric\cite{1109.3706}. $N_r$ represents the
regularized number of zero modes for each field. Computation reviewed in
\cite{1109.3706} shows that $N_r=-1$ for each four dimensional gauge field
and $N_r=-3-K$ for the four dimensional metric
where $K$ denotes the number of rotational isometries of the black hole solution.
The $-3$ in the latter expression comes from the zero modes of the metric
on $AdS_2$ whereas the $-K$ factor comes from the $K$ gauge fields on
$AdS_2$ coming from the dimensional reduction of the metric along the angular
coordinates.\footnote{It may appear strange that the number of zero modes is negative,
but the actual number of zero modes is given by $(1 -\cosh\eta_0)$ times the number
quoted here. Here $\eta_0$ is an infrared cut-off on the coordinate $\eta$. Thus for example
the number of zero modes of the metric on $AdS_2$ is given by $3(\cosh\eta_0-1)$ which
is clearly a positive number. The contribution proportional to $\cosh\eta_0$ can however
be cancelled by boundary counterterms (or equivalently absorbed into a redefinition of the
ground state energy) and only the term proportional to $-3$ contributes to the entropy.
}
Thus for example for the Kerr-Newmann black hole we have $K=1$ and hence
\be \label{ezeromodecounting}
 \sum_r (\beta_r-1)N_r = (1-1)\times (-1) + (2-1) \times (-3-1) = -4\, .
 \ee
The non-zero mode contributions to \refb{elogcoeff} 
have been discussed earlier, {\it e.g.} in \cite{9709064,1104.3712} but without
taking into account the effect of the background gauge fields.

Now for the solution \refb{ekerrmetric} we have\cite{9912320,0302095}
\ben \label{erimric}
R_{\mu\nu\rho\sigma}R^{\mu\nu\rho\sigma}
&=& {8\over (r^2+a^2 \cos^2\psi)^6}
\Big\{ 6 M^2 (r^6 - 15a^2 r^4\cos^2\psi
+ 15 a^4 r^2 \cos^4\psi - a^6 \cos^6\psi) \nn
&& -12 M Q^2 r (r^4 - 10 r^2 a^2 \cos^2\psi + 5 a^4 \cos^4\psi) \nn &&
+ Q^4 (7r^4 - 34 r^2 a^2 \cos^2\psi + 7 a^4 \cos^4\psi)
\Big\} \nn
R_{\mu\nu} R^{\mu\nu} &=& {4 Q^4 \over (r^2 + a^2 \cos^2 \psi)^4} \, .
\een
In the extremal limit we get
\ben \label{eff1}
\int_{\rm horizon} 
 d\psi d\phi \, G(\psi) 
 R_{\mu\nu\rho\sigma}R^{\mu\nu\rho\sigma} &=&
 \frac{8\pi}{b \left(b^2+1\right)^{5/2} \left(2
   b^2+1\right)}\nn &&
 \Bigg\{3   \left(2 b^2+1\right)^2 \tan
   ^{-1}\left(\frac{b}{\sqrt{b^2+1}}\right)\nn &&
   +b \sqrt{b^2+1} \left(-8 b^6-20 b^4
    -8 b^2  +1\right)\Bigg\}\, , \nn 
 \int_{\rm horizon}   d\psi \, d\phi \, G(\psi) \, 
R_{\mu\nu}R^{\mu\nu} &=& \frac{2\pi }{b \left(b^2+1\right)^{5/2}
   \left(2 b^2+1\right)} \nn &&
\Bigg\{3 \left(2 b^2+1\right)^2
   \tan ^{-1}\left(\frac{b}{\sqrt{b^2+1}}\right)
   \nn &&
   +b \sqrt{b^2+1} \left(8 b^2+5  \right)\Bigg\}\, ,
   \een
 where
 \be \label{edefb}
 b = a/Q\, .
 \ee
Note that $b$ remains fixed under the scaling \refb{escaling}.
Using \refb{ea4fin}, \refb{elogcoeff}, \refb{ezeromodecounting},
\refb{eff1}  and the result 
$A_H\sim \Lambda^2$
we now get the logarithmic correction to black hole entropy
to be
\ben \label{elogfin}
-2 \ln A_H - {1\over 720} \, \ln A_H
\frac{1 }{b \left(b^2+1\right)^{5/2}
   \left(2 b^2+1\right)} 
\Bigg\{1233 \left(2 b^2+1\right)^2
   \tan ^{-1}\left(\frac{b}{\sqrt{b^2+1}}\right) \nn
   - b \sqrt{b^2+1} \left(-463 + 3080 b^2 + 7960 b^4 + 3184 b^6\right)\Bigg\}   \, .
   \een
This describes logarithmic correction in the microcanonical ensemble
where $J_3$ is fixed to $J$ but $\vec J^2$ is arbitrary. As discussed in
\cite{1109.3706}, if we also fix $\vec J^2$ to $J(J+1)$ then  in the scaling limit
\refb{escaling} the coefficient of the $\ln A_H$ term does not change.
On the other hand if we consider the Reissner-Nordstrom black hole
and fix both $J_3$ and $\vec J^2$ to 0, then the number $K$ of rotational isometries
changes from 1 to 3, and as a result the  zero mode contribution
changes from $-2\ln A_H$ to $-3\ln A_H$. 

We can get the result for Kerr black hole by taking the $b\to \infty$ limit
of \refb{elogfin}. This gives
\be \label{ekerr}
- 2\ln A_H + {199\over 90} \ln A_H\, ,
\ee
in agreement with the result of \cite{1109.3706} (eq.(3.12) with $n_V=1$,
$n_S=n_F=n_{3/2}=0$).
On the other hand to get the result for the extremal Reissner-Nordstrom
black hole with $\vec J=0$, $\vec J^2=0$ we need to take the
$b\to 0$ limit, and replace the zero mode contribution $-2\ln A_H$ by $-3\ln A_H$ 
since the
number of rotational isometries now goes up to 3. This
gives
\be \label{ern}
-3 \ln A_H - {106\over 45} \ln A_H\, .
\ee
This agrees with the result of \cite{1108.3842} (eq.(4.40)). 

\sectiono{Discussion}

As mentioned earlier, our result \refb{elogfin} puts strong constraint on
any attempt at a microscopic explanation of the entropy of extremal
Kerr-Newmann black holes. Such a theory must reproduce not only the
leading Bekenstein-Hawking term but also the subleading logarithmic
correction given in \refb{elogfin}. In particular if there is an underlying two
dimensional conformal field theory behind the entropy of the extremal
Kerr-Newmann
black hole, as has been suggested in \cite{0809.4266}, then \refb{elogfin} 
could be used as a guideline to search for this
underlying CFT.  In this context we would like to note that \refb{elogfin} is
different from the universal form of the logarithmic correction that appears in
the Cardy limit. This is not necessarily a contradiction to the Kerr/CFT 
hypothesis since the limit involved here corresponds to taking the central charge
as well as the $L_0$ eigenvalue to be large and of the same order\cite{0809.4266}
 whereas
the Cardy limit corresponds to taking the $L_0$ eigenvalue to be large keeping
the central charge fixed. Indeed, even for supersymmetric extremal black holes
for which the macroscopic and the microscopic 
computations of the logarithmic correction to the entropy
match, both the macroscopic and the
microscopic results differ from what one would expect by a  naive application of the
Cardy formula\cite{1109.3706}.

\bigskip

{\bf Acknowledgement:} 
This work was
supported in part by the project 11-R\&D-HRI-5.02-0304
and the J. C. Bose fellowship of 
the Department of Science and Technology, India.

\end{document}